\documentclass[11pt]{article}
\usepackage[]{psfig}
\newcommand{\be}{\begin{equation}}
\newcommand{\ee}{\end{equation}}
\newcommand{\ba}{\begin{eqnarray}}
\newcommand{\ea}{\end{eqnarray}}

\begin{document}

\title{\bf Some Noncommutative Multi-instantons from Vortices in Curved Space}
\author{
D.H.~Correa\thanks{CONICET} \,,
E.F.~Moreno$^*$ and
F.A.~Schaposnik\thanks{Associated with CICPBA}\\
{\normalsize\it  Departamento de F\'\i sica, Universidad Nacional
de La Plata}\\
{\normalsize\it C.C. 67, 1900 La Plata, Argentina}
}

\date{\hfill}
\maketitle

\begin{abstract}
We construct $U(2)$ noncommutative multi-instanton solutions by extending
Witten's ansatz \cite{witten} which reduces the problem of cylindrical
symmetry in four dimensions to that of a set of
Bogomol'nyi equations for an Abelian Higgs model in
two dimensional curved space. Using the Fock space approach, we give
explicit vortex solutions to the Bogomol'nyi equations and, from them,
we present multi-instanton solutions.
\end{abstract}
\date{}

After  the first instanton solution with topological charge $Q=1$ was presented
in \cite{BPST},
many efforts were devoted to the construction of  $Q=n$
instantons, as well as to  the analysis
 of free parameters of the general
solution. A first successful result was reported in
\cite{witten}, where a cylindrically symmetric multi-instanton solution
 was
constructed  by relating the problem
with that of vortex solutions in two-dimensional curved space.
After 't Hooft proposal of a very simple ansatz \cite{tH}, another family
of  multi-instantons
was constructed \cite{jack}. Finally,
a systematic method for finding instanton solutions and their moduli
space, the so called ADHM construction \cite{ADHM}, was developed.

Instantons were rapidly recognized as a basic ingredient for studying
non-perturbative aspects of quantum field theories. They are also
relevant in the context of string theory
and noncommutative geometry. Concerning this last issue,
 instantons in  noncommutative  {\bf R}$^4$ space
were first presented in \cite{ns} where the ADHM
construction was adapted to the noncommutative case. In this
work the discussion was
mainly centred in the $U(1)$ gauge group and $q=1$ instanton. Prompted
by this work, many other applications of the ADHM method in
noncommutative space were presented \cite{SW}-\cite{Par}.

The alternative 't Hooft ansatz approach to noncommutative instantons was also
analysed in \cite{ns}, but
 problems with selfduality were overlooked in that work. These problems
 were overcome in \cite{Correa}, where a regular solution for the $q=1$ U(2)
instanton was explicitly constructed by an appropriate extension
of  the 't Hooft ansatz (related work on this issue was discussed
in \cite{Kim}-\cite{Lec}). Concerning multi-instantons, the
analogous of 't Hooft solution for $U(N)$ instantons with $N>1$
has not
 been found (some problems preventing their  construction were already
  discussed in \cite{Correa}).

 As in the commutative case, there is still the possibility to
 look  for  noncommutative multi-instanton $U(2)$ solutions by connecting the
 problem with that of (noncommutative) vortex solutions in curved space, the
 analogous of the solution presented in \cite{witten} in ordinary space.
 It is the purpose of the present work to analyse this issue which has the
 additional interest of requiring the construction of noncommutative solitons
  in a  nontrivial metric.

~

In order to extend the approach in \cite{witten},
connecting an axially symmetric ansatz for the instanton gauge field
with vortex solutions in 2-dimensional curved space-time, we shall consider
the following commutation relations for cylindrical coordinates
($(r,\vartheta,\varphi)$ and $t$) in Euclidean
4-dimensional space,
\begin{eqnarray}
&& [r,t] = i\theta(r,t)\label{rt}\\
&& [r,\vartheta] = [r,\varphi]=[t,\vartheta] = [t,\varphi] =
[\vartheta,\varphi] = 0
\label{tehtas}
\end{eqnarray}
Eq.(\ref{tehtas}) corresponds to the most natural commutation relations
to impose when a problem with cylindrical symmetry is to be studied. Although
 $\theta(r,t)$ in (\ref{rt}) is in principle an arbitrary function,
 we shall see that, in the reduced 2-dimensional problem,
 a covariantly  constant
$\theta$ guarantees associativity of the noncommutative product of functions.
As we shall see,
this in turn implies
\be
\theta(r,t) = \theta_0 r^2
\label{elec}
\ee
with $\theta_0$ a dimensionless constant. We shall then take (\ref{elec})
as defining noncommutativity of coordinates $r,t$ from here on.

Thanks to the fact that $\theta$ is covariantly constant in
two-dimensional space, there is a change
of coordinates that greatly simplifies calculations.
Indeed,  taking $(r,\vartheta,\varphi,t) \to (u = -1/r,\vartheta,\varphi,t)$,
eq.(\ref{rt}) reduces to
\be
[u,t] = i \theta_0
\label{const}
\ee
while (\ref{tehtas}) remain unchanged (with $r$ replaced by $-1/u$).
Note that in terms of curvilinear coordinates $(u = -1/r,\vartheta,\varphi,t)$,
$\theta^{\mu\nu}$ is constant and hence the Moyal product $*$
provides a realization of the noncommutative product of functions.

We shall take the gauge group to be  $U(2)$ and
define
\be
F_{\mu\nu} = \partial_\mu A_\nu - \partial_\nu A_\mu + i [A_\mu,A_\nu]
\label{FF}
\ee
with
\be
A_\mu = \vec{A}_\mu \cdot \frac{\vec\sigma}{2} + A_\mu^4 \frac{I}{2}
\label{tI}
\ee
and $\vec \sigma =(\sigma^a)$ the Pauli matrices.
The dual field strength $\tilde F_{\mu\nu}$ is defined as
\be
 {\tilde{F}}_{\mu \nu} \equiv \frac{1}{2}
 \, \varepsilon_{\mu\nu\alpha\beta}
{\sqrt g} g^{(\alpha)} g^{(\beta)}F_{\alpha\beta}
\label{var}
\ee
Here we have used that the metric tensor associated with curvilinear
coordinates
$(u = -1/r,\vartheta,\varphi,t)$ is diagonal. Its components read
\be
g^{\mu\nu} = {\rm diag} (u^4,u^2,\frac{u^2}{\sin^2 \vartheta},1) \equiv
\left(g^{(u)},g^{(\vartheta)},g^{(\varphi)},g^{(t)}\right)
 \label{met}
\ee
We shall look for multi-instanton solutions to the
 selfduality equations
\be
F_{\mu\nu} = \pm {\tilde{F}}_{\mu \nu}
\label{sd}
\ee
To this end, we consider an  ansatz for the gauge field components, which
is the
$U(2)$ noncommutative extension of the one solving the commutative case.
 For the $SU(2)$ sector we just take Witten's ansatz \cite{witten},
 which can be written as
\begin{eqnarray}
\vec{A}_u  & = &  A_u(u,t) \vec \Omega(\vartheta,\varphi)  \nonumber\\
\vec{A}_t  & = &  A_t(u,t) \vec \Omega(\vartheta,\varphi) \nonumber\\
\vec A_\vartheta &=& \phi_1(u,t) \partial_\vartheta \vec
\Omega(\vartheta,\varphi) +
\left(1 + \phi_2(u,t)\right) \vec \Omega (\vartheta,\varphi)
 \wedge \partial_\vartheta \vec \Omega (\vartheta,\varphi)
\nonumber\\
\vec A_\varphi &=& \phi_1(u,t) \partial_\varphi \vec \Omega(\vartheta,\varphi)
 +
\left(1 + \phi_2(u,t)\right) \vec
\Omega (\vartheta,\varphi) \wedge \partial_\varphi \vec \Omega
(\vartheta,\varphi)
\label{ansatz}
\end{eqnarray}
with
\begin{equation}
\vec \Omega(\vartheta, \varphi)  = \left(
\begin{array}{c}
\sin \vartheta \cos \varphi \\
\sin \vartheta \sin \varphi \\cos \vartheta
\end{array}
\right)
\ee
Concerning the  remaining $U(1)$ components, it is natural
to propose the ansatz
\begin{eqnarray}
A_u^4 &=& A_u^4(u,t)  \nonumber\\
A_t^4 &=&  A_t^4(u,t) \nonumber\\
A_\vartheta^4 \!\! &=&\!\! A_\varphi^4 = 0
\label{cuatro}
\end{eqnarray}
With this ansatz, the selfduality equations (\ref{sd}) become
\begin{eqnarray}
&&\partial_t A_u - \partial_u A_t + \frac{i}{2} [A_t,A_u^4]
+ \frac{i}{2} [A^4_t,A_u] = 1 - \phi_1^2 - \phi_2^2
\nonumber\\
&&\partial_t A_u^4 - \partial_u A_t^4 + \frac{i}{2} [A_t^4,A_u^4]
+ \frac{i}{2} [A_t,A_u] =  -i[\phi_1,\phi_2]
\nonumber\\
&&\partial_t \phi_1 + \frac{1}{2} [A_t,\phi_2]_+ + \frac{i}{2} [A_t^4,\phi_1]
= u^2\left( \partial_u\phi_2 - \frac{1}{2} [A_u,\phi_1]_+ + \frac{i}{2}
[A_u^x,\phi_2]
\right)
\nonumber\\
&&\partial_t \phi_2 - \frac{1}{2} [A_t,\phi_1]_+ + \frac{i}{2} [A_t^4,\phi_2]
= -u^2\left( \partial_u\phi_1 + \frac{1}{2} [A_u,\phi_2]_+ + \frac{i}{2}
[A_u^x,\phi_1]
\right) \nonumber\\
\label{largas}
\end{eqnarray}
As in other instanton analysis \cite{Correa} one could restrict
even more the ansatz for the $U(1)$ sector so that the form of the equations
for the $SU(2)$ components become the natural noncommutative
generalization of those in ordinary space. We then propose the following
identification
\begin{eqnarray}
A_t^4(u,t) = A_t(u,t) \nonumber
\\
A_u^4(u,t) = A_u(u,t) \label{simpl}
\end{eqnarray}
With this, and introducing the notation
\begin{eqnarray}
\phi &=& \phi_1 - i \phi_2 \nonumber
\\
D\phi &=& \partial \phi +i A\phi \nonumber\\
F_{tu} &=& \partial_t A_u - \partial_u A_t + i[A_t,A_u]
\label{tres}
\end{eqnarray}
system (\ref{largas}) reduces to
\begin{eqnarray}
F_{tu} &=& -\frac{1}{2}[\phi,\bar \phi] \label{suno}\\
F_{tu} &=& 1-\frac{1}{2}[\phi,\bar \phi]_+ \label{sdos}\\
D_t\phi &=& i u^2 D_u\phi\
\label{cortas}
\end{eqnarray}
This system of equations is one
of the main steps in our task of  constructing multi-instantons
 and deserves some comments. It is an overconstrained system but,
 as we shall see, nontrivial solutions can be found.
 The two last equations
resemble the Bogomol'nyi equations arising in ordinary two
dimensional {\it curved} space.
In fact,
they coincide (in the commutative limit) with those discussed in \cite{witten},
with coordinates $u = -1/r, t$ and a   metric
\be
g^{\mu\nu} = \left( \matrix{
u^2  & 0 \cr
0 & {u^{-2}} \cr}
\right)
\label{newmetric}
\ee
with $\det g(x) = 1$.
Hence,  we
succeded in connecting, also in the noncommutative
case, instanton selfdual equations in 4 dimensional Euclidean
space working in curvilinear
coordinates $(u = -1/r,\vartheta,\varphi,t)$
with vortex Bogomol'nyi equations in 2 dimensional curved space with coordinates
$(u,t)$. Were we able to find nonconmmutative multivortex solutions, we
then could
explicitly write noncommutative multi-instanton solutions, as done in
\cite{witten} for the commutative theory.

There is however the third equation (\ref{suno}) in the
coupled system (\ref{suno})-(\ref{cortas}), a
remnant
of the $U(1)$ sector necessarily present
in the noncommutative case. We shall see
however that this equation becomes identical to
eq.(\ref{sdos})  for a particular
family of scalar field solutions. We then pass to analyse this and the
obtention of vortex solutions in curved noncommutative space.

At this point, we have to handle a noncommutative field theory in
curved space. This problem can be related to that of defining a
noncommutative product with $\theta$ depending on the coordinates.
Let us briefly recall this last problem. Consider a general
noncommutative product $\circ$ such that
\be
[y^m,y^n]_\circ = y^m\circ y^n - y^n\circ y^n =
i\theta^{mn}(y)
\label{circ}
\ee
Associativity
of the $\circ$ product can bee seen to impose the following condition
on $\theta^{ab}$ \cite{kontsevich}-\cite{GHS},
\begin{equation}
\theta^{mn} \partial_n \theta^{ab}
+\theta^{an} \partial_n\theta^{bm} +
\theta^{bn} \partial_n \theta^{ma} = 0
\label{once}
\ee
which is equivalent to
\begin{equation}
\theta^{mn} \nabla_n \theta^{ab}
+\theta^{an} \nabla_n \theta^{bm} +
\theta^{bn} \nabla_n \theta^{ma} = 0
\label{doce}
\ee
with
\be
\nabla_m \theta^{ab} = \partial_m\theta^{ab} +
\Gamma^a_{\,m s} \theta^{s b} +
\Gamma^b_{\, ms} \theta^{as}
\label{trece}
\ee
Here $\Gamma^a_{\,m s}$ is the Christoffel symbol associated with the
two dimensional metric $g_{ab}(y)$.
Now, as exploited in \cite{GHS}, a covariantly constant $\theta^{ab}$
\be
\nabla_m\theta^{ab}=0
\label{catorce}
\ee
trivially verifies (\ref{doce}), and hence leads to an associative product.
One can see that, in two dimensions, eq.(\ref{catorce})
reduces to
\be
\frac{1}{\sqrt g} \partial_m \left( \sqrt g
\theta^{ab} \right) = 0
\label{quince}
\ee
where $g = \det g_{ab}$. Then,
the most general covariantly constant $\theta^{ab}$ takes
the form
\be
\theta^{mn} = \frac{\varepsilon^{mn}}{\sqrt g}
\label{dieciseis}
\ee

Let us consider at this point the two-dimensional metric relevant to
the vortex problem in ordinary space. It corresponds, in coordinates
$(r,t)$ to $g^{ab} = r^2 \delta^{ab}$. According to eq.(\ref{dieciseis}),
a covariantly constant
$\theta^{ab} = \varepsilon^{ab} \theta(r,t)$ takes, in such a metric, the form
\be
\theta(r,t) = \theta_0  {r}^2
\label{tetacero}
\ee
This is precisely the form chosen for $\theta(r,t)$
in eqs.(\ref{rt}),(\ref{elec}). Now, in the
  coordinate system $ (u=-1/r,t)$  the $\circ$ commutator
 becomes the ordinary Moyal commutator  since
\be
[u,t]_\circ = i\theta_0
\label{quesi}
\ee

Going back to the problem of
solving the Bogomol'nyi
system (\ref{suno})-(\ref{cortas}), it is convenient   to define
  complex variables
\be
z = \frac{1}{\sqrt {2}} (u + i t)
\, , \;\;\; \;\;\; \bar z = \frac{1}{\sqrt {2}} (u - i t)
\ee
in terms of which eqs.(\ref{suno})-(\ref{cortas}) become
\begin{eqnarray}
 \left( 1 - \frac{1}{2} (z + \bar z )^2\right)D_z \phi &=&
 \left( 1 + \frac{1}{2} (z + \bar z )^2\right)D_{\bar z} \phi
\label{cuss} \\
iF_{z \bar z} & = & 1 - \frac{1}{2}[\phi,\bar \phi]_+
\label{cus}\\
iF_{  z \bar z} & = & - \frac{1}{2}[\phi,\bar \phi] \label{cu}
\end{eqnarray}
In view of equation (\ref{quesi}),
if one defines
\begin{equation}
z \to \sqrt{\theta_0}\; a \;  \;\;\;
\bar z   \to  \sqrt{\theta_0}\; a^\dagger
\end{equation}
one then has
\begin{equation}
[a,a^\dagger]  = 1
\end{equation}
and hence one is lead to follow the alternative Fock space
approach
 to noncommutative field theories,
 taking $a$ and
$a^\dagger$ as annihilation and creation operators generating the
Fock space $\{|n\rangle\}$.
Concerning derivatives, one has
\be
\partial_z \to  -\frac{1}{\sqrt{\theta_0}}\;[ a^\dagger,~]
\, , \;\;\;\; \partial_{\bar z} \to \frac{1}{\sqrt{\theta_0}}\; [a,~]
\label{conmutads}
\ee

One should notice at this point that,
in the  case at hand, the complex variable $z$ is defined in the half-lower
semiplane and this could cause problems when connecting the product
of operators in Fock space with Moyal products in coordinate representation.
In fact,  this connection
  can be
established through an isomorphism which results in a mapping
between operators in  Fock space and functions in $R^{2n}$.
If the two-dimensional
manifold is not $R^2$ but a half plane   one should analyse
whether  the  isomorphism is  modified. Instead,   we shall follow
an alternative approach which consist in doubling the space manifold
so as to work in $R^2$ and exploit the ordinary connection. Afterwards,
we shall restrict the solutions to the
relevant domain. Having in mind the features of Witten's solutions in ordinary
space, with an even magnetic
field associated     as a
function of $u= -1/r$, we shall   seek for solutions with such a magnetic
field behavior in $R^2$.

 Note that eq.(\ref{cus}) coincides with the corresponding
\underline{flat space} original Bogomol'nyi equation for the magnetic field.
It is the equation (\ref{cuss})
governing the scalar field dynamics where the curved space metric
plays a r\^ole. As discussed before, there is also the new third equation
 (\ref{cu}) arising from the   additional $U(1)$ sector.

One can easily see that compatibility of (\ref{cus}) and (\ref{cu}),
 implies
\be
\bar \phi \phi = 1
\ee
and hence the only kind of nontrivial solutions following our ansatz
should have the form
\be
\phi = \sum_{n=0} |n + q\rangle \langle   n|
\ee
With this, it is easy now to construct a class of solutions analogous
to those found in
\cite{Poly}-\cite{Bak} for noncommutative Nielsen-Olesen vortices in
flat space. Indeed, take
\begin{eqnarray}
\phi &=& \sum_{n=0}|n+q\rangle\langle n|  \nonumber\\
A_z &=& -\frac{i}{\sqrt{\theta_0}} \sum_{n=0}^{q-1} \left( \sqrt{n+1}
\right) |n+1\rangle\langle n|\label{solucion2} +\nonumber\\
&& ~ ~ ~ +\frac{i}{\sqrt{\theta_0}} \sum_{n=q} \left( \sqrt{n+1-q} - \sqrt{n+1}
\right) |n+1\rangle\langle n|
\label{solucion}
\end{eqnarray}
One can trivially verify
 that functions (\ref{solucion}) satisfy eqs.(\ref{cuss})-(\ref{cu}) provided
 $\theta_0 = 2$. In
 particular, both the l.h.s. and r.h.s of eq.(\ref{cuss}) vanish separately.
 Regarding the particular value of $\theta_0$ for which we find a solution,
 let us recall that also for vortices in  flat space it was necessary to
 fix $\theta_0 = 1$ in order to satisfy the corresponding Bogomol'nyi
 equations\cite{JKL}-\cite{Bak}.

 The magnetic field $B_{\theta_0} = i F_{z\bar z}$
  associated with
 solution (\ref{solucion2}) takes the form,
\be
B_{\theta_0 = 2} = -\frac{1}{2} \left(|0\rangle\langle 0|+ ...
+ |q-1\rangle\langle q-1|\right) \equiv B
\label{bb}
\ee
with associated magnetic flux

\be
\Phi = \pi \theta_0 Tr B_{\theta_0} = -{\pi} q
\ee
Note  the factor $\pi \theta_0$ in the definition of the magnetic flux. It
is one half of the usual factor, since our actual
problem  corresponds to the half plane.

We can now easily write  the self-dual multi-instanton solution in
$4$-dimen\-sio\-nal
space by inserting the solution (\ref{solucion}) in ansatz (\ref{ansatz}). The
resulting selfdual field strength reads
\begin{eqnarray}
\vec F_{tu} &=& B \vec \Omega \nonumber\\
\vec F_{\vartheta\varphi}
&=& B   \sin \vartheta \, \vec \Omega \nonumber\\
F_{tu}^4 &=& B \nonumber\\
 F_{\vartheta\varphi}^4 &=& B \sin \vartheta
\label{instan}
\end{eqnarray}
with the other field-strength components vanishing.
The instanton number is then given by
\begin{equation}
Q = \frac{1}{32 \pi^2} {\rm tr} \int d^4 x {\varepsilon}^{\mu \nu \alpha \beta}
 F_{\mu\nu} F_{\alpha \beta}
= \frac{1}{\pi}
\int_{-\infty}^{0}du \int_{-\infty}^{\infty} dt B^2 = 2 {\rm Tr} B^2
= \frac{q}{2}
\end{equation}
We thus see that $Q$ can be in principle integer or semi-integer,
 and this for ansatz (\ref{ansatz})
which is formally the same as that proposed in \cite{witten} for ordinary
space and which yielded in that case to an integer. The origin of this
difference between the commutative and the noncommutative cases can be traced
back to the
fact that in the former case,  boundary conditions imposed on the half-plane
force the solution to have an associated integer number. We were not
able to find  boundary conditions such that semi-integer
 configurations
were excluded. Then, in order to have regular instanton solutions with integer
number we  just restrict vortex configurations  (\ref{solucion}) to those
 with $q = 2n$
so that the instanton number $Q = n \in Z$. In fact, if one plots Witten's
 vortex solution in ordinary space in
the whole $(u,t)$ plane, the magnetic flux has two peaks and the corresponding
vortex number is even.  Our choice then corresponds, in Fock space,
 to selecting the analogous of that solution in noncommutative space.

 In summary, we have presented multi-instanton solutions to the
 $U(2)$ selfduality equations (\ref{sd}) in a noncommutative space
 where the commutation relations for coordinates are those
 defined by eqs.(\ref{rt})-(\ref{tehtas}). In such a noncommutative space
 the problem can be reduced to that of solving Bogomol'nyi equations
 for vortices in curved space. Using the Fock space approach
 one can easily find a
 family of  solutions for the latter,
 eqs.(\ref{solucion})-(\ref{bb}), and, from them, to explicitly construct
 instanton solutions, (\ref{instan}), with  arbitrary charge. Let us end
 by recalling that  axially symmetric  instantons in
 ordinary 4-dimensional space-time can be employed to seek for
 multimonopole solutions in 3-dimensional space
 \cite{manton}-\cite{Ch}. One could then follow a similar
 approach in the noncommutative case in the search
 of explicit magnetic  monopole solutions. We hope to discuss this issue
 in a future work.

\vspace{1 cm}

\noindent\underline{Acknowledgements}:
 We wish  to acknowledge Peter Forg\'acs for collaboration
 in the first stages of this work.
 We also thank R.~Schiappa for useful correspondence. This work  is partially
 supported
 by UNLP, CICBA,  ANPCYT,  Argentina.
E.F.M. are partially supported by Fundaci\'on Antorchas, Argentina.



\begin{thebibliography}{99}
\bibitem{witten} E.~Witten, Phys. Rev. Lett. {\bf 38} (1977) 121.
\bibitem{BPST} A.A.~Belavin, A.M.~Polyakov, A.S.~Schwarz and Yu.S.
Tyupkin, Phys. Lett. {\bf B59} (1975) 85.
\bibitem{tH} 't Hooft, 1976, unpublished.
\bibitem{jack} C.~Nohl, R.~Jackiw and C.~Rebbi, Phys. Rev. {\bf D15} (1977)
1642.
\bibitem{ADHM}M.~F.~Atiyah, N.~J.~Hitchin, V.~G.~Drinfeld and Y.~I.~Manin,
Phys.\ Lett.\ A {\bf 65} (1978) 185.
\bibitem{ns} N.~Nekrasov and A.~Schwarz,
Commun.\ Math.\ Phys.\  {\bf 198} (1998) 689.
\bibitem{SW} N.~Seiberg and E.~Witten, JHEP {\bf 09} (1999) 032.
\bibitem{fki} K.~Furuuchi,
  Prog.~Theor.~Phys.{\bf  103} (2000)
1043;
Commun.\ Math.\ Phys.\  {\bf 217}, 579 (2001); hep-th/0010006;
   JHEP {\bf 0103}, 033 (2001),
\bibitem{su} A.~Schwarz,
Commun.\ Math.\ Phys.\  {\bf 221}, 433 (2001).
\bibitem{Par}
S.~Parvizi,
Mod.\ Phys.\ Lett.\ A {\bf 17}, 341 (2002).
\bibitem{Correa}
D.~H.~Correa, G.~S.~Lozano, E.~F.~Moreno and F.~A.~Schaposnik,
Phys.\ Lett.\ B {\bf 515} (2001) 206.
\bibitem{Kim}
K.~Y.~Kim, B.~H.~Lee and H.~S.~Yang,
Phys.\ Lett.\ B {\bf 523}, 357 (2001)
\bibitem{Lec}
O.~Lechtenfeld and A.~D.~Popov,
JHEP {\bf 0203}, 040 (2002).
 \bibitem{kontsevich} M.~Kontsevich, q-alg/9709040.
\bibitem{cataneofel} A.S.~Cattaneo and G.~Felder,
Commun.Math.Phys. {\bf 212} (2000) 591; 
 L.~Cornalba, JHEP {\bf 0009} (2000)017,
 L.~Cornalba and R.~Schiappa,
Commun.Math.Phys. {\bf 225} (2002) 33.
S.~Cacciatori, A.~H.~Chamseddine, D.~Klemm,
L.~Martucci, W.~A.~Sabra and D.~Zanon,
hep-th/0203038.
\bibitem{GHS} G.~Gibbons, C.~Herdeiro and G.~Silva, unpublished.
\bibitem{Poly}A.~P.~Polychronakos,
Phys.\ Lett.\ B {\bf 495} (2000) 407.
\bibitem{JMW} D.P.~Jaktar, G.~Mandal and S.R.~Wadia, JHEP {\bf 0009}
 (2000) 018.
\bibitem{JKL} J.A.~Harvey, P.~Kraus and F.~Larsen,
JHEP {\bf 0012}  (2000) 024.
\bibitem{Bak}
D.~Bak, Phys.\ Lett.\ B {\bf 495} (2000) 251.
\bibitem{manton} N.S.~Manton, Nucl. Phys. {\bf B135} (1978) 319.
\bibitem{forgacsmanton} P.~Forg\'acs and N.S.~Manotn,
Commun. Math. Phys. {\bf 72} (1980) 15.
\bibitem{fhp} P.~Forg\'acs, Z.~Horvath and L.~Palla,
Nucl. Phys. {\bf B221} (1983) 235.
\bibitem{Ch} A.~Chakrabarti Nucl. Phys. {\bf B248} (1984) 2019.
\end{thebibliography}
\end{document}